# Liquified protein vibrations, classification and cross-paradigm *de novo* image generation using deep neural networks

Markus J. Buehler[a,*]

[a] Laboratory for Atomistic and Molecular Mechanics (LAMM), Massachusetts Institute of Technology, Cambridge, MA, United States of America

* Correspondence: E-mail : mbuehler@MIT.EDU, Phone: +1.617.452.2750

**Abstract:** In recent work we reported the vibrational spectrum of more than 100,000 known protein structures, and a self-consistent sonification method to render the spectrum in the audible range of frequencies (Extreme Mechanics Letters, 2019). Here we present a method to transform these molecular vibrations into materialized vibrations of thin water films using acoustic actuators, leading to complex patterns of surface waves, and using the resulting macroscopic images in further processing using deep convolutional neural networks. Specifically, the patterns of water surface waves for each protein structure is used to build training sets for neural networks, aimed to classify and further process the patterns. Once trained, the neural network model is capable of discerning different proteins solely by analyzing the macroscopic surface wave patterns in the water film. Not only can the method distinguish different types of proteins (e.g. alpha-helix vs hybrids of alpha-helices and beta-sheets), but it is also capable of determining different folding states of the same protein, or the binding events of proteins to ligands. Using the DeepDream algorithm, instances of key features of the deep neural network can be made visible in a range of images, allowing us to explore the inner workings of protein surface wave patter neural networks, as well as the creation of new images by finding and highlighting features of protein molecular spectra in a range of photographic input. The integration of the water-focused realization of cymatics, combined with neural networks and especially generative methods, offer a new direction to realize materiomusical "Inceptionism" as a possible direction in nano-inspired art. The method could have applications for detecting different protein structures, the effect of mutations, or uses in medical imaging and diagnostics, with broad impact in nano-to-macro transitions.

## 1. Introduction

Proteins are the basic building blocks of life. They form materials as diverse silk, cells, and hair, but also offering other functions from enzymes to drugs, and pathogens like viruses [1–6]. While we cannot see small nanoscopic objects like proteins or other molecules, a common feature is their continuous motion, or vibration, that can be understood as an overlay of fundamental normal modes each consisting of harmonic waves. In recent work [1] we reported the vibrational spectrum of more than 100,000 known protein structures, and a self-consistent sonification method to render their complex vibrational spectrum in the audible range of frequencies. The sonification work has also been used to train generative neural networks to facilitate the design of *de novo* proteins using machine learning [2,7].

This article focuses on a different perspective and reports a distinct, complementary and translational approach, in which we transform these molecular vibrations into vibrations of thin water films using acoustic actuators, leading to visual images of complex materialized patterns of surface waves. This approach follows the pioneering concept developed by Chladni [8] and is similar to the cymatics method [9–14], offers a distinct means to assess the molecular details of protein vibrations in the form of macroscopic vibrations visible to our eyes. In addition to potential applications in outreach, the physical manifestation of molecular vibrations at the macroscale provides a novel avenue to render sound visible, providing an alternative method to conventional musical notation [11] and novel interactive approaches to interact with sound using senses other than our ears [15–18]. Moreover, the use of artificial intelligence provides an exciting avenue to classify, process and understand, or augment images generated by sound. We will explore some of these augmentation concepts in this paper, by highlighting key features of images as detected by distinct layers in deep neural networks. These connections between sound, materialization and images can offer many avenues for future research, enabled by significant progress made in recent years in computer vision [2,19–25].

Figure 1A shows the overall flow of the research reported here. The approach includes the calculation of molecular vibrational spectra transposed to audible frequencies and made audible (see earlier work for details, [1,6]), which are then used to excite a thin film of water. Images of surface wave patterns are collected, building a training set for a machine learning model. The predictive power of the classifier model is validated, showing that the model can correctly determine the protein structure solely based on images of the surface wave patterns. Figure 1B depicts an overview of eight distinct protein structures (chosen to reflect different complexity and size in its molecular structure) that are investigated in this paper.

## 2. Materials and Methods

Sonification is a method to translate data structures into audible signals, which has been explored widely as a means to better understand scientific data in a range of areas of application including spider webs, proteins, and other systems [26–34]. Here, we utilize the protein synthesizer published in earlier work [1] and generate various audio signals that are fed into an actuator attached to a petri dish with a thin layer of water.

### 2.1 Protein structure labeling

All protein codes are expressed as Protein Data Bank (PDB) [35] identifier, and can be downloaded and further explored at www.rcsb.org.

### 2.2 Experimental Setup

The experimental setup is shown in Figure 2. The system consists of an actuator attached to a petri dish. The actuator (Dayton Audio DAEX25 Sound Exciter, driven by an analog amplifier) is driven by an amplifier, who receives the analog audio signal directly from a digital-to-analog (D/A) audio interface in a laptop computer. A 2,100 lumens LED light source is used for illumination during image capturing (Winplus, LED Folding Worklight).

The actuator is fixed permanently to a solid and immobile substrate with high mass. The petri dish has been spray-painted white for a clear optical signal (unless indicated otherwise; as we generated some images with a reflective aluminum foil background for better contrast). Water is inserted into the petri dish with a height of 1 mm. The higher the water level, the more actuation energy is needed, so we pick a level that provides with significant surface wave generation at the available actuation energy. A camera is mounted at a fixed location for consistent images.

### 2.3 Sound generation

The digital audio workstation (DAW) Ableton Live [36] and Max/MSP [37] patches, as described in earlier work [1], are utilized here to generate the audio signals; these provide implementations of the protein synthesizer. A few comparative simple, pure sine wave forms are generated as well.

**2.4 Image capturing and pre-processing**

Images are taken in the form of a continuous video, recorded at 60 fps, and with 1920x1080 resolution, recorded with a Samsung Galaxy S10+ 5G camera (the camera combines a primary 12 Mp camera and variable-aperture lens with a 16 Mp ultra-wide-angle component). The video is stored, and subsequently individual frames are extracted, using a Python code. The images are cropped consistently so as to focus on the surface wave structures inside the petri dish, without showing a boundary. Each image has a size of 585x788 after such processing, and before being fed into the neural network. No additional image processing is conducted. Figure 3B shows an example of the surface wave structures that emerges in the experiment, as used in the deep neural network training.

The majority of images are used for training (80%), and 20% for validation. To test the prediction of the model, new data from additional imaging of the experiment is used.

**2.5 Neural network design and training**

Two neural networks are considered in this work.

First, an existing ResNet neural network model as reported in [38] that has been pre-trained against a large dataset (https://tfhub.dev/tensorflow/resnet_50/feature_vector/1). This TensorFlow Hub model uses the implementation of ResNet with 50 layers, and contains a pre-trained instance of the network, arranged to get feature vectors from images. We use transfer learning to retrain all parameters in the neural network, by adding a dense layer after the feature layer to add a new set of classifications to reflect the distinct audio sources of molecular spectra. The ResNet model features 23,587,789 parameters [38].

Second, a custom designed neural network with much fewer parameters (consisting of alternating convolutional and pooling layers), as a comparative approach. Figure 4 depicts a summary of this neural network design. The model consists of multiple sequences of convolutional and pooling layers, following a standard image classifier design. The custom designed neural network features a total of 945,165 parameters, and 20 layers.

**2.6 DeepDream generative method**

We use DeepDream [39] to generate novel images by activating select layers in the deep neural network, based on the model trained against the water surface images. This method is an approach to visualize the patterns learned by a neural network, and results in novel images. It can be understood similar to the process of an interpretation of structures in clouds, for instance, in which we try to interpret shapes and forms from the abstract patterns. In the case of the neural network, patterns it sees in a given image are interpreted and then accentuated. The method works by processing an image through a neural network, then calculating the gradient of the image with respect to the activations of a set of layers. The image is then adapted to enhance the patterns seen by the neural network.

In processing the images, we explore different layers for the activation algorithm, each resulting in distinct features being highlighted in the image (e.g. internal convolutional layer vs. internal pooling layers). The purpose of this computation is to explore the inner structures and features learned by the neural network, and make them visible.

**3. Results**

We collect a number of images, around 1,000 for each case, and feed the labeled images into the neural network training process. A total number of 13 labels are used, reflecting 13 distinct frequency spectra corresponding to the individual protein structures considered here. The audio signals of a consistent, stationary spectrum corresponding to the molecular vibrational signature of each of these structures are fed into the actuator. The cases included in the training set are:

- Flat: No actuation (i.e., no audio signal)
- Pure sine waves: 3 distinct frequencies (Sine33 = 50 Hz, Sine43 = 98 Hz, Sine65 = 349 Hz)
- Various protein structures, including protein data bank IDs: 6vsb, 1o7m, 5xdj, 6m17, 6m18 (and others). Figure 1B depicts an overview of all proteins considered in this study, where the first row represents relatively simple structures (added complexity from left (short alpha-helix) to right (more complex alpha-helical fold). The lower row features more complex protein structures, including two spike proteins from the pathogen of COVID-19 (two distinct molecular states - 6vxx and 6vsb of COVID-19), as well as the COVID-19 spike protein bound (6m17) and unbound (6m18) to the human ACE2 receptor.

Figure 5 depicts the results of the training process for both neural network formulations. Both cases show reasonable convergence. The pre-trained ResNet case (Figure 5A) converges faster and in a more stable manner, however. This makes sense given that the model starts from pre-trained parameters.

Figure 6 shows results of a classification test suite conducted with the two trained models, depicting averaged values for a series of unknown, new images, fed to the neural network. The radar plot shows the score for each of the 13 labels (each is denoted out of 1.00). The data shows that both models predict the case very well, and without error. It can also be seen that the score for the correct prediction is quite pronounced, showing that the classifier works very well and can distinctly classify the correct protein structure at the origin of the acoustic actuation. This can be verified by the fact that the highest score in the plot is close to 1.00, and that there is a distinct peak for each of the cases.

Going into more detail, we analyze a few specific cases. For instance, the orange color represents the scores for protein 1akg. The highest score, in both models, is for the label 1akg, which indicates correct classification. For 3tnu, the highest score is 3tnu, and so on. Notably, both models are capable of predicting the correct labels, solely from images of surface waves, for all cases studied. It is remarkable that the method works for a variety of protein structures with distinct frequency spectra from complex to simple including the pure sine wave cases at different frequencies. It also renders correct predictions for proteins with distinct geometric dimensions. For instance, 1akg is a very small protein that features only a simple fold and a short alpha-helix segment. 3tnu is a small alpha-helix fragment without any folding. In contrast, 6vsb is a very large protein. Other interesting cases are 6m17 and 6m18. The protein 6m17 shows only minute distinctions from 6m18 in terms of its vibrational spectrum, as is shown in Figure 7. Yet, the model can distinguish the cases very well.

Figure 8 depicts an application of DeepDream [39] to generate novel images by activating select layers in the deep neural network, based on the model trained against the water surface images. This approach offers a new way to design generative art from molecular vibrations. The method can be applied to various detailed aspects, including an analysis of images of surface waves from the experiment, or the analysis of other, distinct images. The method can be viewed as a means of reflections of molecular vibrational pattern identification in other image sources. The examples depicted here show how the method can accentuate photographs of water, but also entirely different objects and landscapes. It can be seen that when fed images of the type of the original images of thin water layers vibrating, the model recognizes the patterns and overlays these quite well on top of it (Figure 8, left column). In the other examples, the fit is not as good, even though the model still captures relevant features in these images.

Figure 9 shows a collection of examples of a variety of source images processed, visualizing features seen by the internal layers of the deep neural network, by selecting different layers. This set of computational experiments visualizes to what extent the features of surface wave patterns are recognized in images of other types. It can be recognized how distinct layers of the neural network capture certain features in the images, at different scales. Figure 10 shows a similar experiment, this time applied to the same image, but considering the result for using individual layers to generate features. The collection of examples of images visualize features seen by the internal layers of the deep neural network, by selecting different layers. Figure 11 shows an analysis of images with DeepDream, using the Inception deep neural network [19]. Unlike in the cases discussed above, here the neural network model used is taken "as trained" against a large number of conventional images (and has not been trained against the dataset reported here). It is clear that the conventional image classification model features very distinct patterns, reflecting the characteristic features learned in a conventional image classifier. It offers interesting insights into what image features are seen by a model like Inception in the abstract representation of surface wave patterns induced by proteins.

## 4. Discussion and conclusion

The new experimental-computational method reported here provides a new way to analyze and visualize molecular vibrations expressed in sound. While the method could in principle be used to predict visuals of conventional audio signals (e.g. specific musical instruments, songs, varied classes of music, etc.) the focus in this paper was on determining the protein structure solely from the pattern of surface waves. While cymatics, or the general attempt to make sound into images is a concept explored since ancient times (since the original works in the late 1700s [8,9]), here we report for the first time a method to visualize molecular vibrations in protein structures in that way, and offer a path to integrate it with artificial intelligence methods. The integration of the water-focused rendering of cymatics, combined with neural networks and especially generative methods, offer a new direction to realize materiomusical "Inceptionism" [39] as a possible future direction in nano-inspired art.

Our method shows that even minute changes in the molecular vibrational spectra (e.g. 6m17 vs 6m18, as shown in Figure 6) can be detected through this method, as well as different states of proteins (e.g. different molecular states of the COVID-19 virus spike protein, 6vxx vs 6vsb). Since we now have the vibrational spectra of all known protein structures available [1], the approach reported here could potentially be used to build a very large dataset to render visual macroscopic representations of all known proteins. While this is beyond the scope of the present study, which focused on the establishment of the basic methodology, such nano-to-macro translations could find applications in a variety of fields. Further work could also be done by providing detailed fluid-structure-interaction models of how the specific patterns emerge. Such research may provide a theoretical basis to complement the experimental research done here.

In terms of limitations, the method reported here can only classify protein structures that the neural network has been trained against. However, future work could seek to generalize the method. Other limitations are the dependence of the images on the experimental setup. It is likely that the patterns of surface waves depend on the particular geometries of the experimental setup, such as the water level, the size of the contained, the power output of the actuators and so on. More broadly, looking ahead, the method can find applications as a means to accentuate images, find patterns in other biomaterials, or to use the approach in detecting patterns across manifestations of domains.

**Acknowledgements:** This work was supported by MIT CAST via a grant from the Mellon Foundation, with additional support from ONR (grant # N00014-16-1-2333) and NIH U01 EB014976.

**Figures and captions**

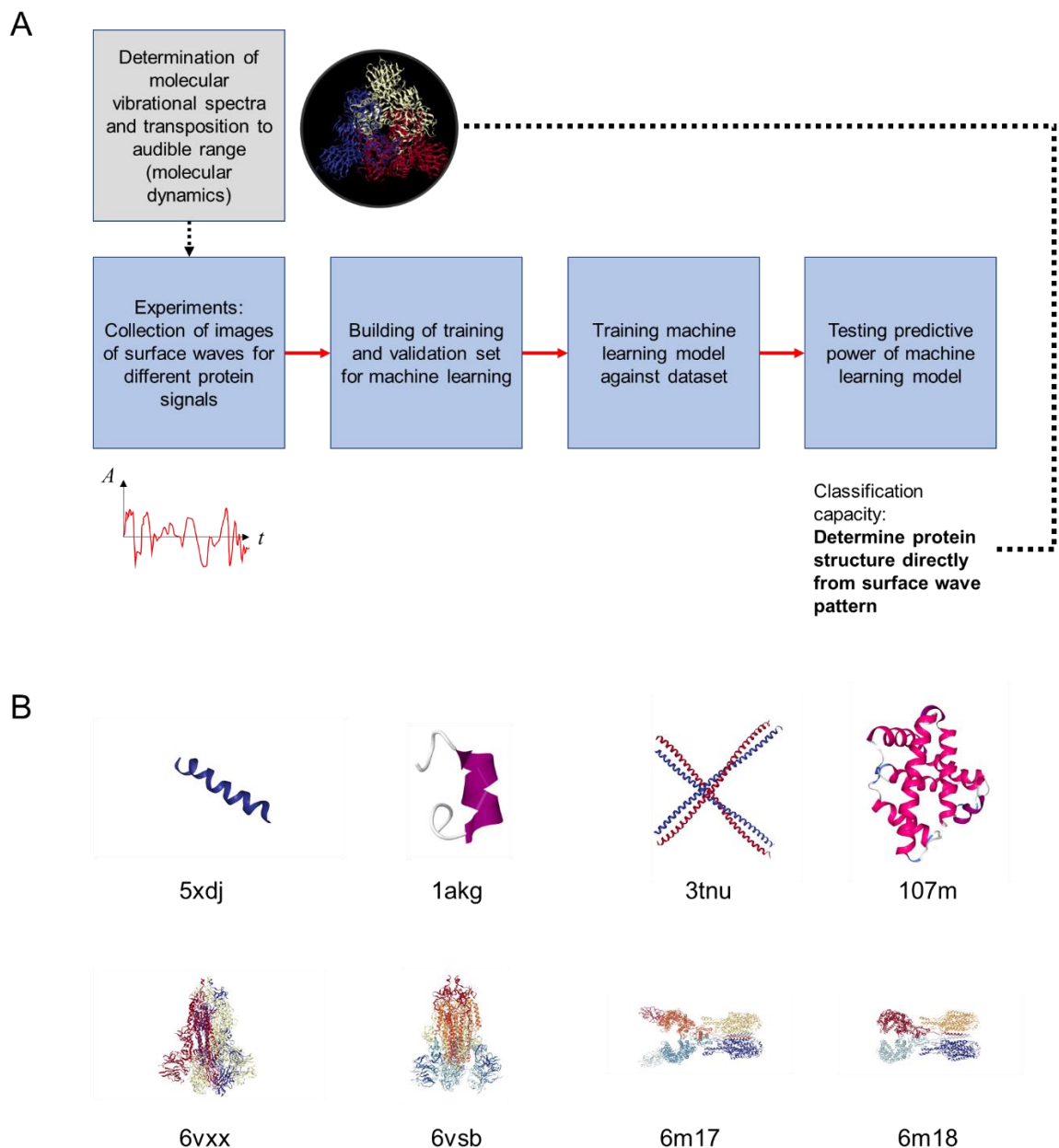

**Figure 1:** A, Overall flow of this research, ranging from the molecular vibrational spectra transposed to audible frequencies, and used to excite a thin film of water using an actuator. Images of surface wave patterns are collected, building a training set for a machine learning model. The predictive power of the classifier model is validated, showing that the model can correctly determine the protein structure solely based on images of the surface wave patterns. B, overview of the various proteins studied in this paper. The first row represents relatively simple structures (added complexity from left (short alpha-helix) to right (more complex alpha-helical fold). The lower row features more complex protein structures, including two spike proteins from coronaviruses (6vxx and 6vsb of COVID-19), as well as the COVID-19 spike protein bound (6m17) and unbound (6m18) to the human ACE2 receptor.

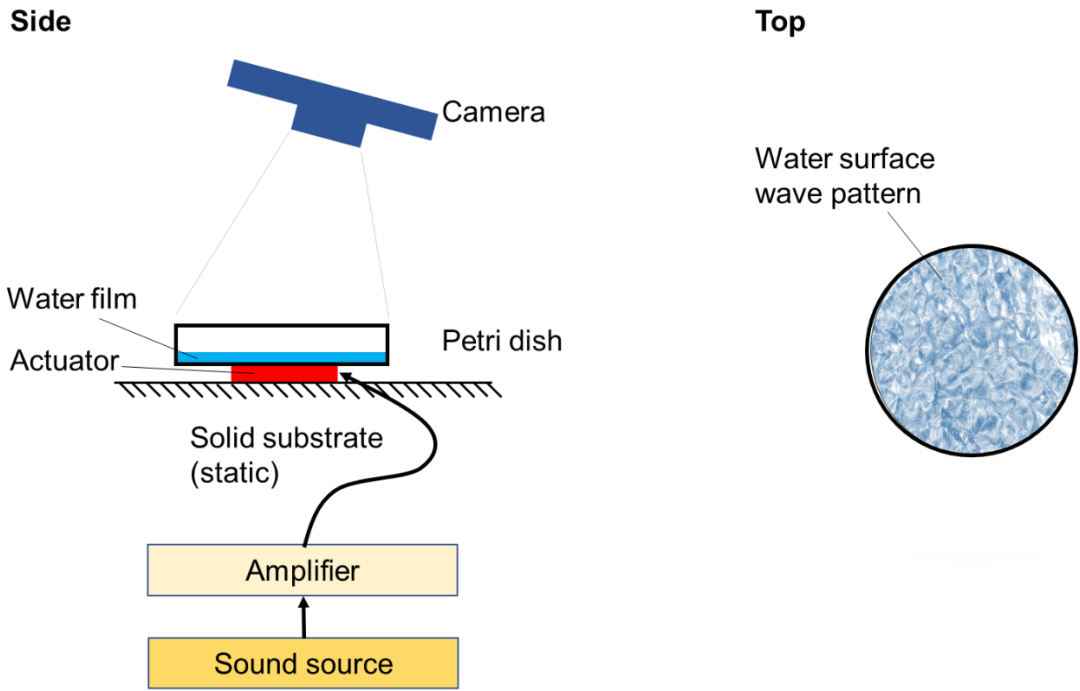

**Figure 2:** Experimental setup used in this study. A think water film in a petri dish is excited by an actuator, resulting in surface waves, as can be seen in the right column labelled "Top". The resulting structures are captured using a high-resolution camera, and collected as a training set for machine learning.

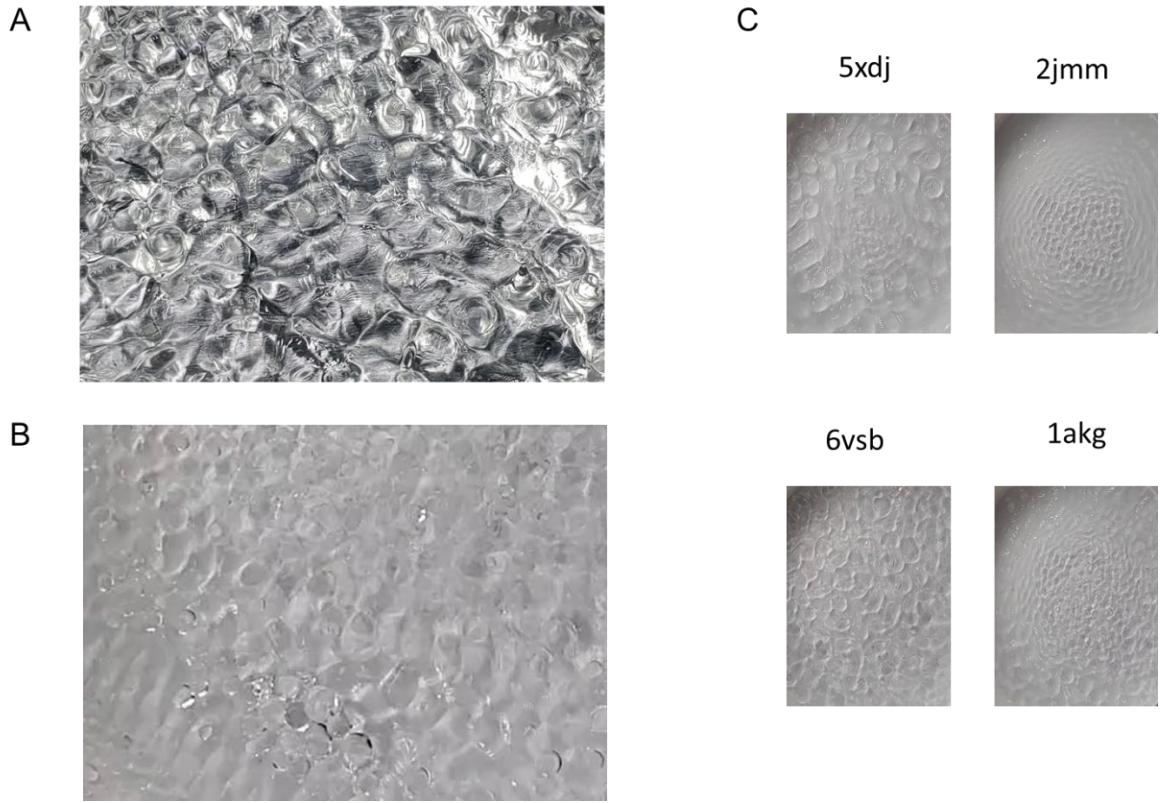

**Figure 3**: Examples of structures of think water film surface wave patterns. A, image taken against a reflective background of aluminum foil. B, Images taken against a solid white background. Images as shown in panel C are used in the neural network training. C, Sample images as used in the training set, in cropped format as fed to the neural network.

```
Layer (type)                 Output Shape              Param #
=================================================================
conv2d_92 (Conv2D)           (None, 510, 510, 32)      896
_________________________________________________________________
max_pooling2d_93 (MaxPooling (None, 255, 255, 32)      0
_________________________________________________________________
batch_normalization_92 (Batc (None, 255, 255, 32)      128
_________________________________________________________________
conv2d_93 (Conv2D)           (None, 253, 253, 64)      18496
_________________________________________________________________
max_pooling2d_94 (MaxPooling (None, 126, 126, 64)      0
_________________________________________________________________
batch_normalization_93 (Batc (None, 126, 126, 64)      256
_________________________________________________________________
conv2d_94 (Conv2D)           (None, 124, 124, 64)      36928
_________________________________________________________________
max_pooling2d_95 (MaxPooling (None, 62, 62, 64)        0
_________________________________________________________________
batch_normalization_94 (Batc (None, 62, 62, 64)        256
_________________________________________________________________
conv2d_95 (Conv2D)           (None, 60, 60, 96)        55392
_________________________________________________________________
max_pooling2d_96 (MaxPooling (None, 30, 30, 96)        0
_________________________________________________________________
batch_normalization_95 (Batc (None, 30, 30, 96)        384
_________________________________________________________________
conv2d_96 (Conv2D)           (None, 28, 28, 32)        27680
_________________________________________________________________
max_pooling2d_97 (MaxPooling (None, 14, 14, 32)        0
_________________________________________________________________
batch_normalization_96 (Batc (None, 14, 14, 32)        128
_________________________________________________________________
dropout_34 (Dropout)         (None, 14, 14, 32)        0
_________________________________________________________________
flatten_17 (Flatten)         (None, 6272)              0
_________________________________________________________________
dense_34 (Dense)             (None, 128)               802944
_________________________________________________________________
dropout_35 (Dropout)         (None, 128)               0
_________________________________________________________________
dense_35 (Dense)             (None, 13)                1677
=================================================================
```

**Figure 4:** Design of the neural network, featuring a total of 945,165 parameters, consisting of alternating layers (there is a total of 20 layers). The model consists of multiple sequences of convolutional and pooling layers, following a standard image classifier design. For comparison, the ResNet architecture features 23,587,789 parameters (with a total of 50 layers) [38].

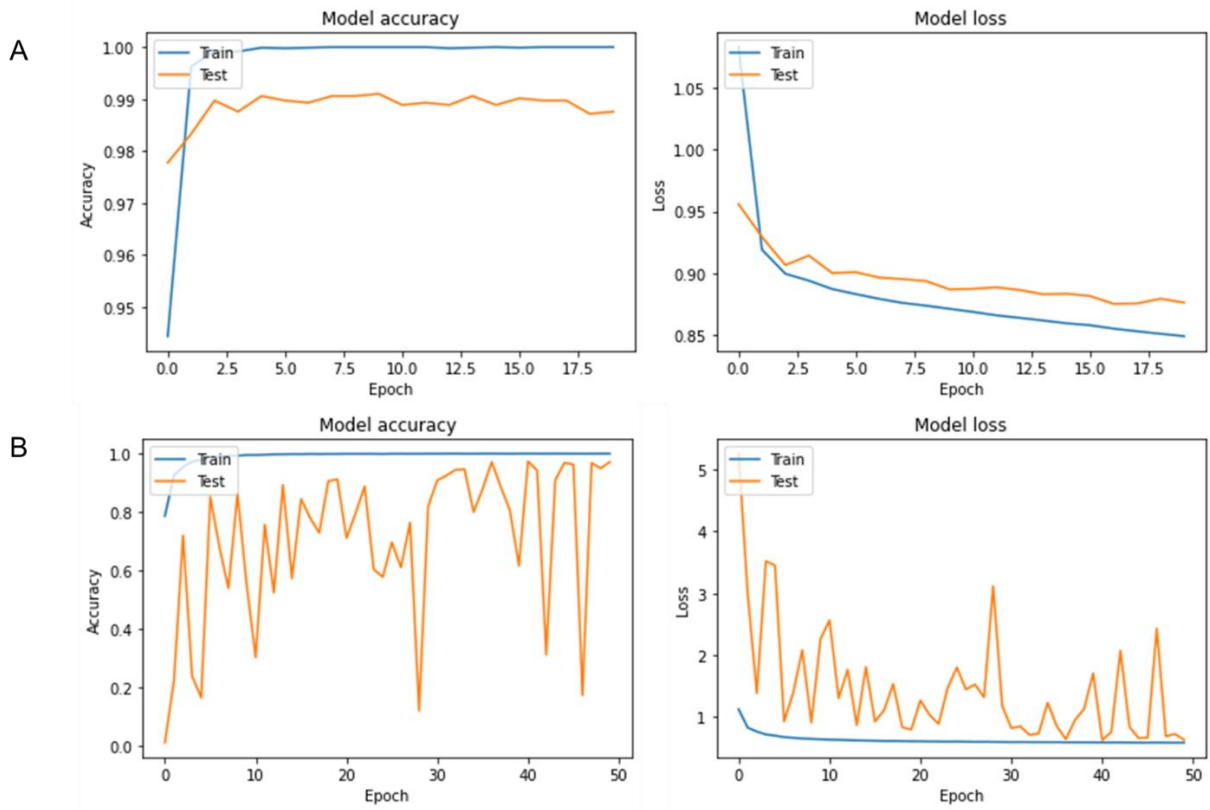

**Figure 5:** Results of the training of the two machine learning models, showing good performance for both cases. A, Training a ResNet model using transfer learning. B, Training a simpler custom neural network design from scratch, whose model structure is summarized in Figure 4.

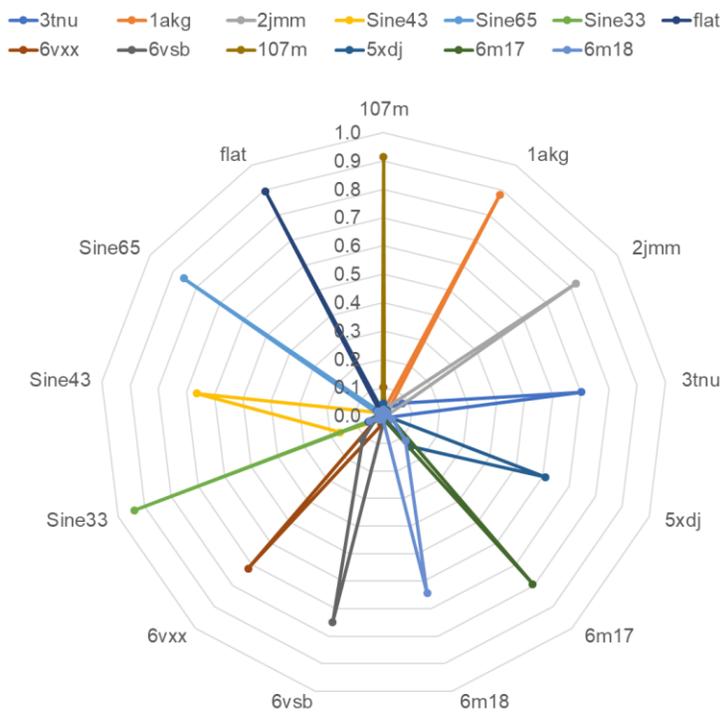

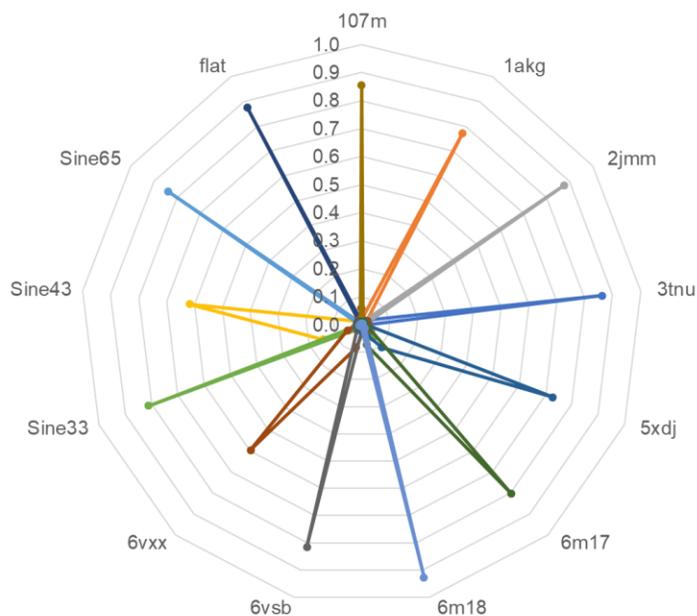

**Figure 6:** Results of the classifier as applied to a new data set that wasn't included in training or validation. The data is plotted in a radar plot (1.0 = the neural network is capable of correctly labeling the image, 0.0 reflects worst performance). We plot averaged scores, providing average values over classifications for multiple images. The data shows that both neural networks are capable of identifying different protein structures, as well as the test cases of sine waves (at different frequencies), as well as silence. For instance, the orange color represents the scores for 1akg. The highest score, in both models, is for the label 1akg, which indicates correct classification. For 3tnu, the highest score is 3tnu, and so on. Both models are capable of predicting the correct labels, solely from images of surface waves, for all cases studied.

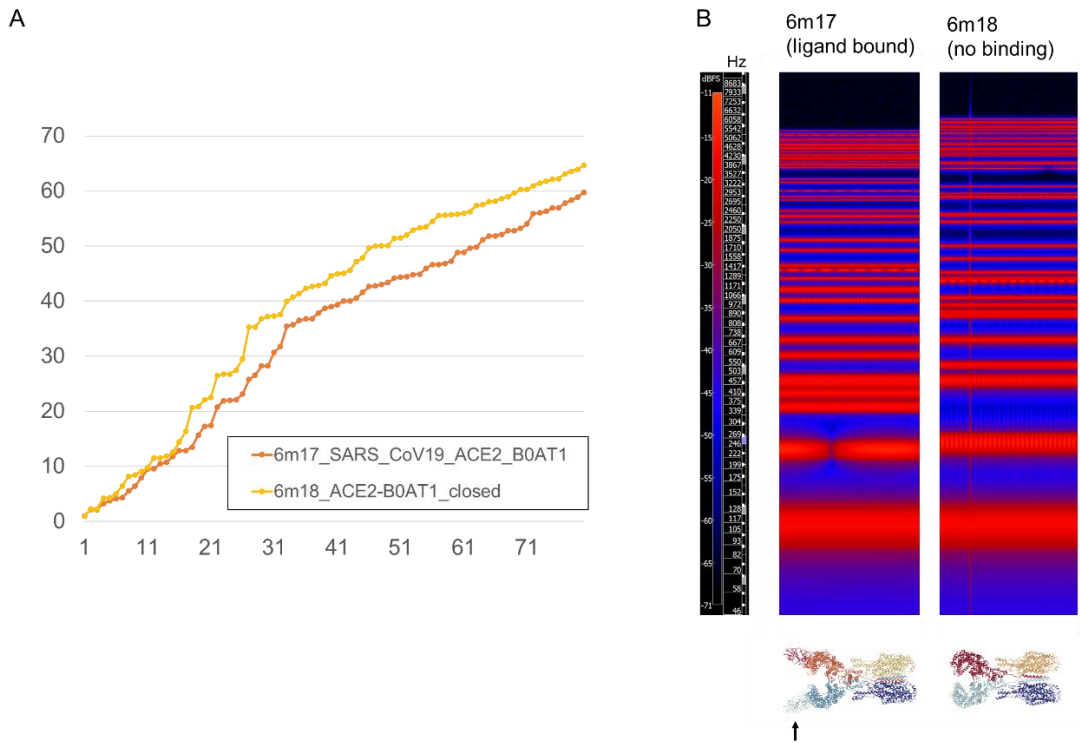

**Figure 7:** Distinct frequency spectrum of normal mode frequencies of proteins with PDB ID 6m17 (virus spike ligand bound, indicated with the arrow) vs. 6m18 (no ligand bound). In spite of the small differences in the structure and the frequency spectrum as depicted in the figure, the machine learning model is capable of easily distinguishing the patterns generated from these audio signals. A, normal mode frequencies of 6m17 and 6m18 over the mode number. B, Spectrogram of the resulting audio signal, using the method reported in [1].

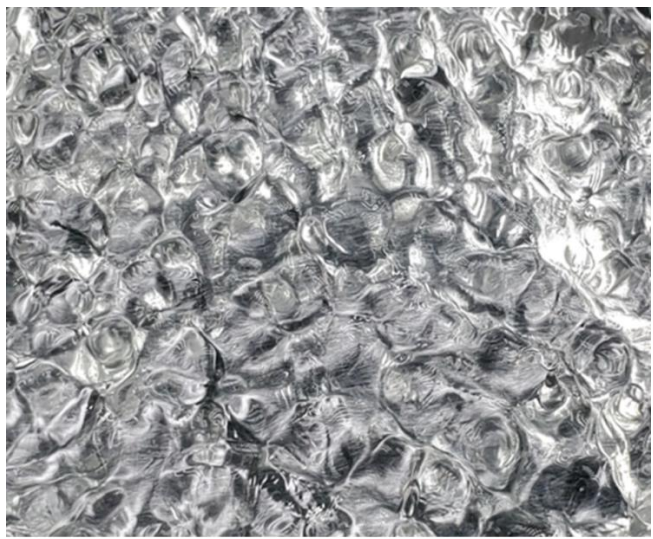
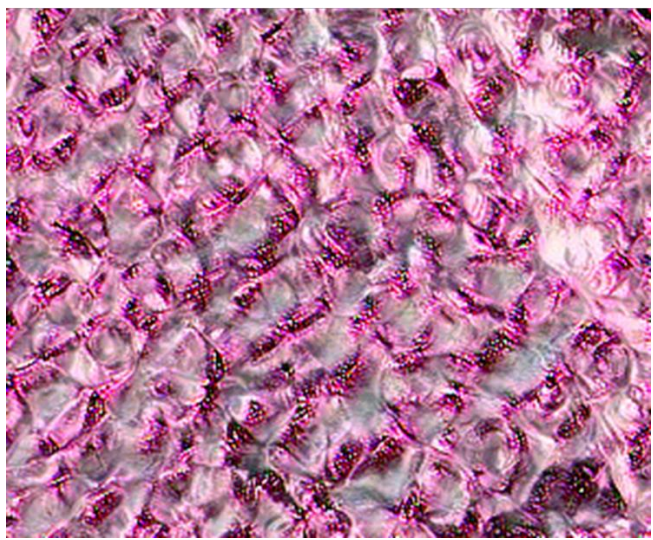

**Figure 8:** An application of DeepDream [39] to generate novel images by activating select layers in the deep neural network, based on the model trained against the water surface images. This approach offers a new way to design generative art from molecular vibrations. The top image is the original, the bottom the processed image. This example shows how the method can be a powerful method to better understand features seen and learned in the neural network, which can be used to accentuate key elements in images.

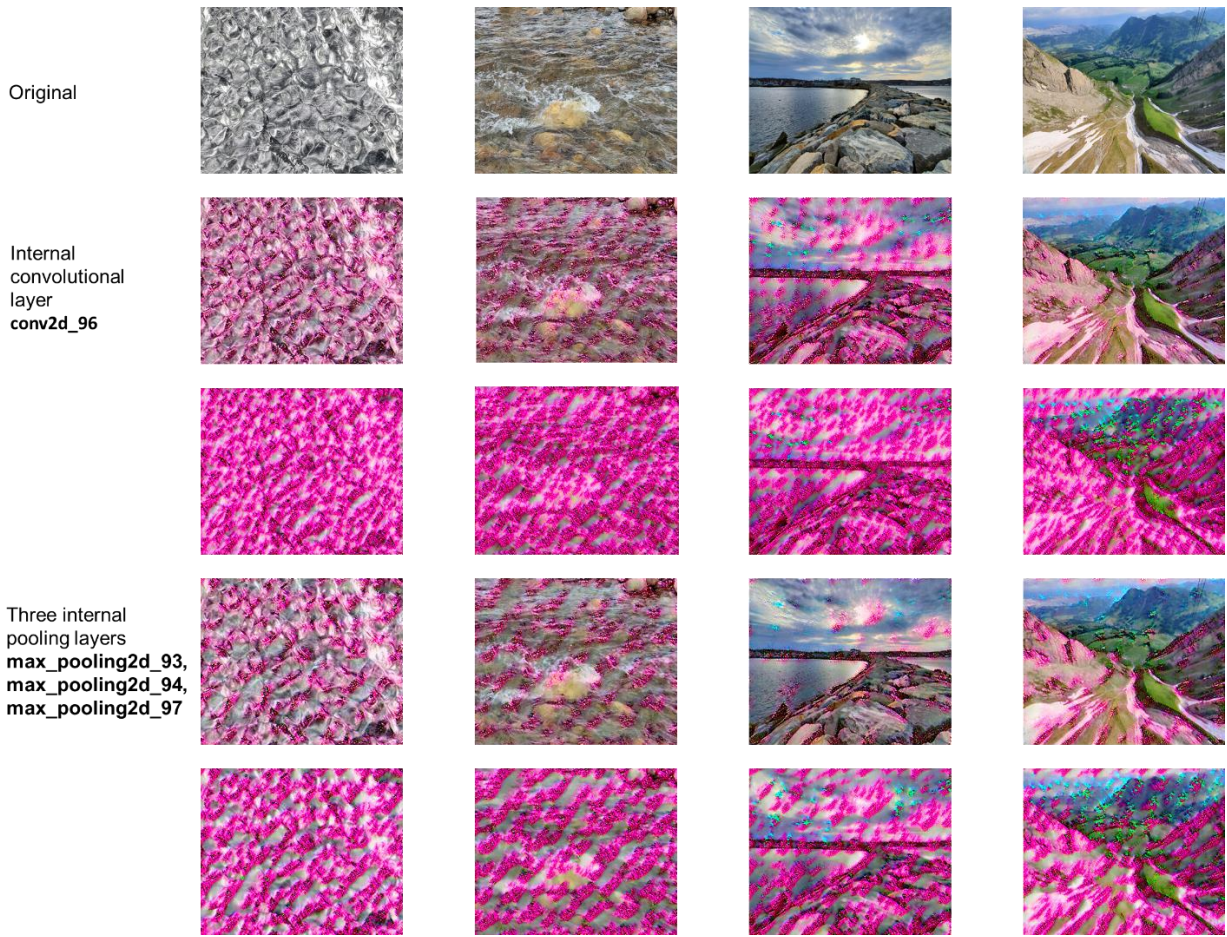

**Figure 9:** Collection of examples of images processed using DeepDream, visualizing features seen by the internal layers of the deep neural network, by selecting different layers. The layers are labeled as reported in Figure 4. It can be recognized how distinct layers of the neural network capture certain features in the images, at different scales.

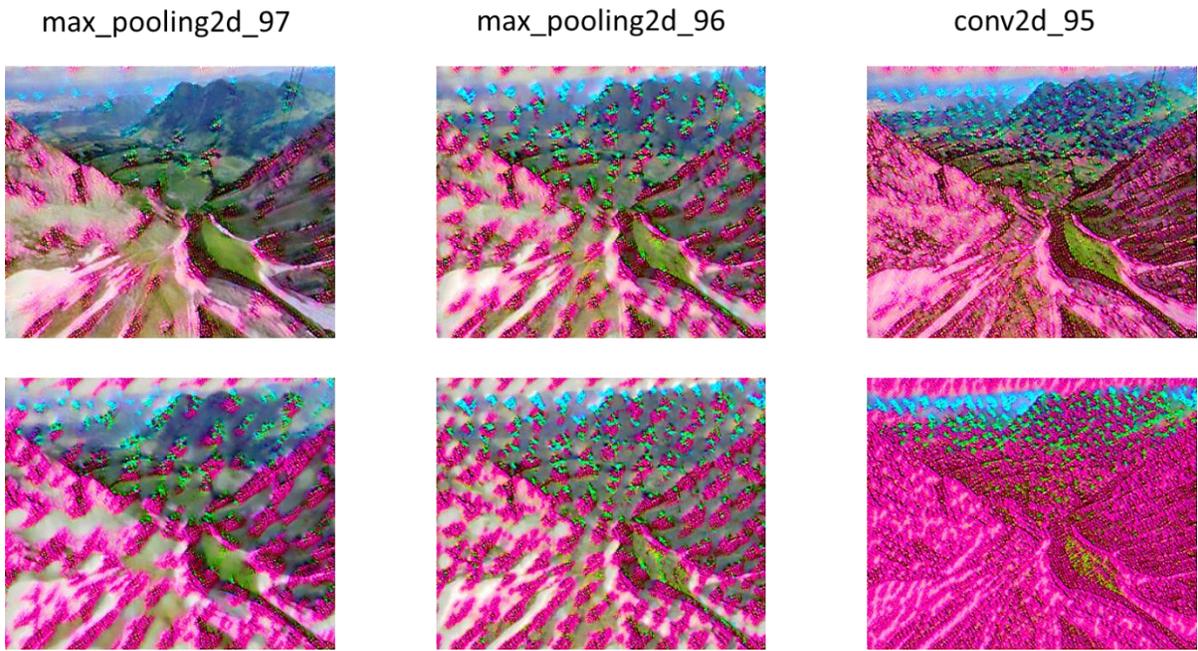

**Figure 10:** Collection of examples of images processed using DeepDream, visualizing features seen by the internal layers of the deep neural network, by selecting different layers. The layers are labeled as reported in Figure 4.

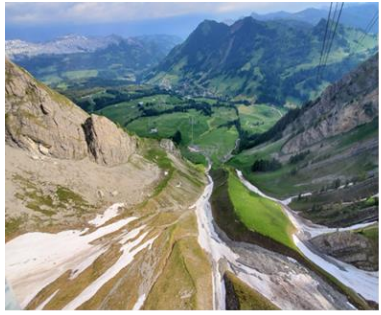 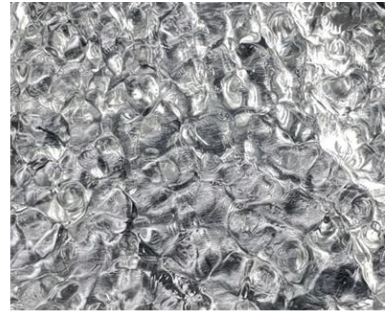
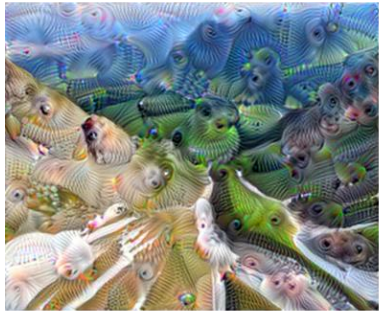 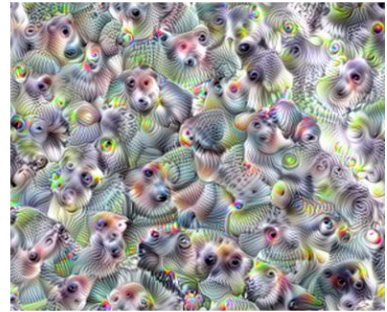
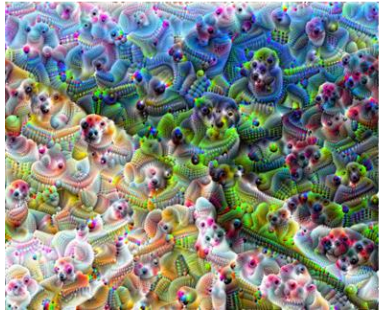 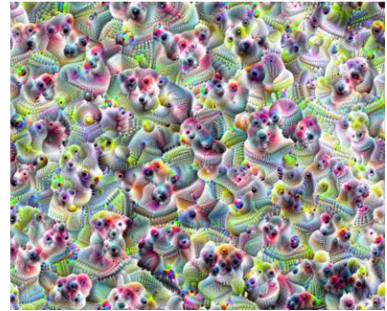

**Figure 11:** Analysis of images with DeepDream, using the Inception deep neural network [19] (note, this model is taken "as is", and has not been trained against the dataset reported here). It is clear that the conventional image classification model features very distinct patterns, reflecting the characteristic features learned in a conventional image classifier. The left column shows the processing resulting from a mountain landscape (same as in Figure 9, right column) for comparison. The right column shows the water surface wave pattern as source image.

19